\documentclass[12pt]{article}
\usepackage{aas_macros}
\usepackage{times}
\usepackage{geometry}
\usepackage[utf8]{inputenc}
\usepackage{enumitem,amssymb}
\usepackage{ragged2e}
\usepackage{graphicx}
\usepackage{fancyhdr}
\usepackage{hyperref}
\usepackage{tabularx}
\usepackage[
    natbib=true,
    style=numeric,
    sorting=none
]{biblatex}
\usepackage{color, colortbl}
\newlist{thematic}{itemize}{8}
\setlist[thematic]{label=$\square$}
\usepackage{pifont}

\newcommand\farcs{\mbox{$.\!\!^{\prime\prime}$}}%

\begin{document}

%% Page 1: Title, technical details, observing description, 1 page limit for this section

\raggedright
\huge{
Demonstrating Improved Contrast on the Roman Coronagraph with Spatial Linear Dark Field Control}
\linebreak
\large

%Thayne Currie (University of Texas-San Antonio/Subaru)
%Olivier Guyon (Subaru)
%Ruslan Belikov (NASA-Ames Research Center)
%Dan Sirbu (NASA-Ames Research Center)
%Mona El Morsy (University of Texas-San Antonio
Thayne Currie$^{1,2}$,
Olivier Guyon$^{2}$, 
Ruslan Belikov$^{3}$, 
Dan Sirbu$^{3}$, 
Mona El Morsy $^{1}$

1 Department of Physics and Astronomy, University of Texas-San Antonio, San Antonio, TX, USA \\
2 Subaru Telescope, National Astronomical Observatory of Japan, 
Hilo, HI, USA\\
3 NASA-Ames Research Center, Moffet Field, CA, USA

\justify{
\textbf{Abstract:}  
The baseline contrast floor from the Roman Coronagraph's High-Order Wavefront Sensing and Control strategy likely degrades over the course of time, requiring periodic recalibration of the dark hole.  Here, we propose to consider spatial linear dark field control (sLDFC) on a one-sided deep-contrast region of the focal plane as a potential test.  Implementing sLDFC on CGI will likely require some unique data acquisition strategies given the EMCCD's high flux sensitivity in long exposures/high gain: we outline three possible approaches.  However, if successful, sLDFC's advances are substantial: (1) enabling us to maintain a fainter, more temporally correlated dark hole which will improve CGI's contrast after post-processing and efficiency (2) providing a reliable signal (bright field) for accurate reconstruction of residual starlight in the dark field, further boosting CGI's detection limit for bright targets.}

\pagebreak
\noindent \textbf{Type of observation:} \\$\boxtimes$   Technology Demonstration\\
$\square$ Scientific Exploration\\\\

\noindent \textbf{Scientific / Technical Keywords:}  
high contrast performance
%\textcolor{blue}{Please choose one or more keywords from the following list:
%disk, companion (exoplanet), companion (substellar), self-luminous, reflected light, other science, observing strategy, post-processing, control algorithm, high contrast performance} \linebreak

\noindent \textbf{Required Detection Limit:}  
%\textcolor{blue}{Please estimate the detection limit necessary for your observation.
%Note: the higher your contrast requirement, the less likely the observation will be feasible in the baseline observing period.  Observations that require detection limits better than 10$^{-8}$ may not be possible.  Observations requiring detection limits from 10$^{-5}$ to 10$^{-6}$ may be appropriate for ``first-look” Coronagraph observations early in the mission, prior to the Coronagraph achieving its full high-contrast performance. }
%\linebreak
%\begin{center}
\begin{tabular}{| c | c | c | c | c |}
\hline
$\geq$10$^{-5}$ & 10$^{-6}$ & 10$^{-7}$ & 10$^{-8}$ & 10$^{-9}$ \\ \hline
 & & x &x & \\ \hline
\end{tabular}
%\end{center}

% \noindent \textbf{Estimate of Observation Length (hours):}  
% %Please estimate the number of hours of observations (including overheads) to achieve your scientific and technical goals).
% \linebreak

% \noindent \textbf{Estimate of Risk Level:}  
% %Please state whether the observation is low, medium, or high risk.
% \linebreak

%Please choose from one of the following supporting or best effort modes (delete or comment out rows from table that you will not use).  %NB: you can select non-supported modes as well, but these modes will likely not be attempted during "first-look" / commissioning and may not be available.

\vspace{0.5cm}
\textbf{Roman Coronagraph Observing Mode(s):} 
%\textcolor{blue}{Please indicate your desired observing mode(s). You may choose from one of the following supported or best effort modes (delete or comment out rows from table that you will not use). You may also choose non-coronagraphic imaging in any of the four filters. You can edit the table to request non-supported modes as well, but these modes will likely not be attempted during the initial ``first-look"/commissioning period and may not be available during the baseline observation phase.  The CPP still solicits observing ideas in non-supported modes as part of a potential further extended observing period with the Roman Coronagraph.}

%\linebreak 
\begin{tabular}{| c | c | c | c | c |}
\hline
Band &   Mode & Mask Type & Coverage & Support \\ \hline \hline

%% Band 1, narrow field imaging and polarimetry
1, 575 nm &   Narrow FoV & Hybrid Lyot & 360$^{\circ}$ & Required (Imaging) \\
 & Imaging &  &  &  \\ \hline 

%% Band 1, wide field imaging and polarimetry
%1, 575 nm &  Wide FoV & Shaped Pupil & 360$^{\circ}$ & Best Effort  \\ 
% &  Imaging &  &  & (Imaging and polarimetry) \\
%\hline

%% Band 2, spectroscopy
%2, 659 nm &  Slit + R$\sim$50  & Shaped Pupil & 2x65$^{\circ}$ & Best Effort \\ 
% &    Prism Spectroscopy &  &  & \\  \hline

%% Band 3, spectroscopy
%3, 730 nm &  Slit + R$\sim$50  & Shaped Pupil & 2x65$^{\circ}$ & Best Effort \\ 
% &    Prism Spectroscopy &  &  & \\  \hline

%% Band 4, wide field imaging and polarimetry
4, 825 nm & Wide FoV  & Shaped Pupil & 360$^{\circ}$ & Best Effort  \\ 
 & Imaging &  & &  (Imaging, Backup Option)  \\ \hline
\end{tabular}

%\pagebreak

%%% Page 2: Table of Example Targets, optional technical questions 

\justify{
%\noindent \textbf{Table of Example Targets:} \textcolor{blue}{Please provide details on one or more example targets.  Target descriptions can remain generic -- e.g. if you just need a non-variable star with V = 5, then it is fine just to provide that detail but not provide a specific star.  Note: the higher your contrast requirement, the less likely the observation will be feasible during the baseline observation period.  Observations requiring detection limits better than 10$^{-8}$ may not be possible -- whether such observations are possible will be determined during coronagraph calibration expected as part of the baseline observation phase.  Target stars should in general have $V$ $\leq$ 5, although observations of somewhat fainter stars down to $V$ $\sim$ 7 may be possible with degraded detection limit performance.}
\begin{center}
\setlength{\tabcolsep}{5pt}
\begin{tabular}{| c | c | c | c |l|}

%\begin{tabular*}{0.95\textwidth}{|p{0.075\textwidth}|p{0.3\textwidth}|p{0.075\textwidth}|p{0.4\textwidth}|}
%$\begin{tabularx}{0.85\textwidth}{|p{0.075\textwidth}|p{0.3\textwidth}|p{0.075\textwidth}|p{0.075\textwidth}|}
\hline
Name &  host star & detection & separation (") & description of target \\
  & V mag. & limit & (or extent)  & \\ \hline \hline
  a very bright star  & $<$3 & 10$^{-8}$ & 0\farcs{}15--0\farcs{}45 & bright PSF star for\\
   &  &  & &EFC/DM probing\\\hline
  a slightly fainter star & 3--5 & 10$^{-8}$  & 0\farcs{}15--0\farcs{}45& target star \\
  %51 Eri b & 5.4 & 7$\times$10$^{-8}$ & 0.23 & self-luminous exoplanet\\
  %HR 8799 e & 5.4 & 7$\times$10$^{-8}$ & 0.23 & self-luminous exoplanet\\
 % AF Lep b & 6.3 & 1$\times$10$^{-8}$ & 0.23 & self-luminous exoplanet\\
  %HD 63754 B & 6.5 & 7$\times$10$^{-8}$ & 0.48 & self-luminous exoplanet\\ % to 2$\times$10$^{-7}$
  %HIP 54515 b & 6.8 & 3$\times$10$^{-6}$ & 0.23 & self-luminous exoplanet\\
  %HD 206893 B & 6.7 & 5$\times$10$^{-9}$ & 0.20 & self-luminous brown dwarf/exoplanet\\
  \hline
  %\footnote{blah}
\end{tabular}
\end{center}

}

% Optional Technical Questions regarding example targets.
%\justify{
%\noindent \textbf{Optional Questions:}
%\textcolor{blue}{Please complete only if you have specific targets in mind -- fine to skip this if you are proposing for a generic category of targets.}
%}

%\justify{
%\noindent \textbf{Are any example targets binary systems in the Washington Double Star survey or other binary survey?} 
%
%}

%\justify{
%\noindent \textbf{Do any of your example Hybrid Lyot coronagraphic target stars have angular diameters $>$ 2 mas?}
%No
%}

\pagebreak

%%% Page 3: Description of Scientific / Technological Goals, 1-page limit for this section

\justify{
% 1 page of Scientific / Technical Justification, including an estimate of time needed.
\begin{figure}[!ht]
    \includegraphics[width=0.9\textwidth,clip]{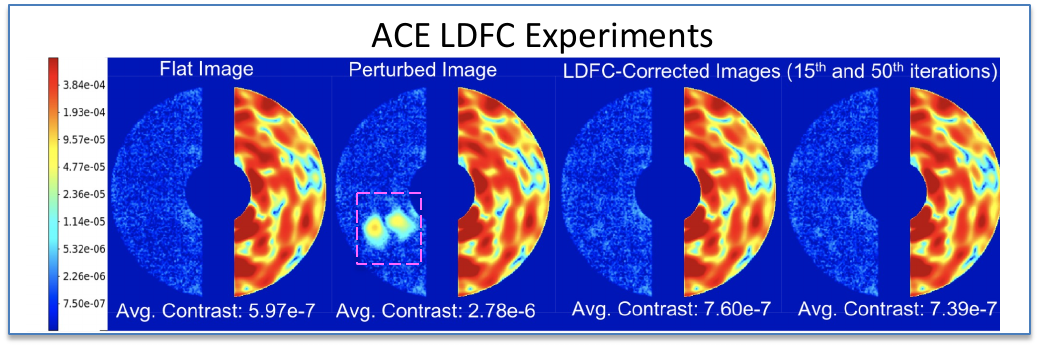}

    \vspace{-0.1in}
    \caption{Spatial Linear Dark Field Control demonstration with the ACE Laboratory in air \citep{Currie2020}.   After an initial DH generated by DM probing, we introduce perturbations that degrade the average DH contrast by a factor of 5.  Spatial LDFC restores the DH average intensity to within 25\% of its original value and sustains this correction.
    }
\end{figure}
\textbf{\Large Anticipated Technology / Science Objectives:}
Imaging planets in reflected light with instruments like the Roman Coronagraphic Instrument (CGI) requires deformable mirror (DM) probing (e.g. electric field conjugation; EFC) to generate a $<$10$^{-8}$-contrast dark hole (DH) within several $\lambda$/D on a point-spread function (PSF) reference and then applying this modulated DM shape to a target star  \citep[][]{Currie2023b,Kenworthy2025}.   However, the DH contrast floor on the target star likely will degrade in average intensity -- and its residual speckles decorrelate -- due to dynamic aberrations inherent in any telescope optical system.  The CGI strategy to mitigate this degradation is to revisit the PSF reference star periodically to recalibrate the dark field before resuming science target exposures, though the actual recovered contrast depth and temporal correlation of residual DH speckles will likely only be known in flight.   Ideally, a wavefront sensing and control (WFSC) approach obviating the need to recalibrate the DH on a PSF reference and leaving more static residuals would increase CGI's efficiency and contrast after post-processing.

Here, we suggest a test of Spatial Linear Dark Field Control (sLDFC) \citep{Miller2017} to maintain the DH without recalibration on a PSF reference star (Fig. 1).  The strategy departs from the standard CGI WFSC strategy in several ways.   After generating a one-sided DH from DM probing methods, the sLDFC loop is commenced, sensing linear changes in the focal plane image region without a DH (i.e. ``the bright field") and reshaping the DM to maintain a static DH, via a simple in-flight response matrix.   The bright field signal serves as input to a PSF reconstruction algorithm aimed at recovering, with high accuracy, the residual starlight component in the DH. This second step has been demonstrated at JPL to boost detection limits by $>$10x, and is being further developed\citep{Guyon2024} (Fig. 2).

Spatial LDFC has been demonstrated on sky behind an extreme AO system and at the Ames Coronagraph Experiment (ACE) Laboratory high-contrast imaging testbed \citep{Miller2017,Ahn2023,Bos2021,Currie2020}.  The ACE experiments (conducted in air, not vacuum) demonstrated a $\approx$5$\times$10$^{-7}$-contrast DH sustained with sLDFC, within $\approx$20\% of the initial DH contrast, with a high temporal correlation, and without evidence for significant null space.

%\footnote{SAT grant "Robust Deep Contrast Imaging with Self-Calibrating Coronagraph Systems", Guyon et al.)}.

%\textbf{Is this observation appropriate for ``first-look" / commissioning ($<$3 months after launch), the observation phase ($<$ 18 months after ``first-look" / commissioning), or a potential extended observing phase?}:

The best time to conduct this experiment is after TTR5 and other coronagraph Objectives are fully achieved but prior to the consideration of any extended mission. A successful sLDFC experiment may allow a more optimal WFSC approach for some science cases.
%This program seeks to fulfill TTR5 contrast requirements (5-$\sigma$ contrast at 575 nm of 10$^{-7}$ or better).  The companion is in an edge-on orbit, moving closer in projected separation with time. Thus, this observation is best executed during the commissioning phase or as early as during the observation phase.
%\textcolor{blue}{Please estimate when in the timeline for the mission your observations would best fit.  This will be dependent on required detection limit. Observations that require only moderate (10$^{-5}$ to 10$^{-6}$) detection limits are especially valuable for the early phases of observations, including ``first-look" observations.  Observations that require stringent detection limits or particularly long observations will not be possible until a potential extended observing phase.  It would be very useful to the CPP if you can indicate whether the proposed observations are low contrast and appropriate for a ``first-look" image or achievable during commissioning, moderate contrast achievable during baseline observations, or require detection limits or long observations that may only be possible during potential extended observations.}

%We propose HLC/575 nm observations of a newly-discovered brown dwarf from the Subaru/OASIS survey of young accelerating stars, which is funded by NASA to provide CGI/TTR5 targets.   The target lies close to a candidate PSF reference that may be within the continuous viewing zone for Roman.  A high SNR detection of this companion would singlehandedly fulfill TTR5 as well as at least one key science Objective.
}

\begin{figure}[!ht]
    \includegraphics[width=0.9\textwidth,clip]{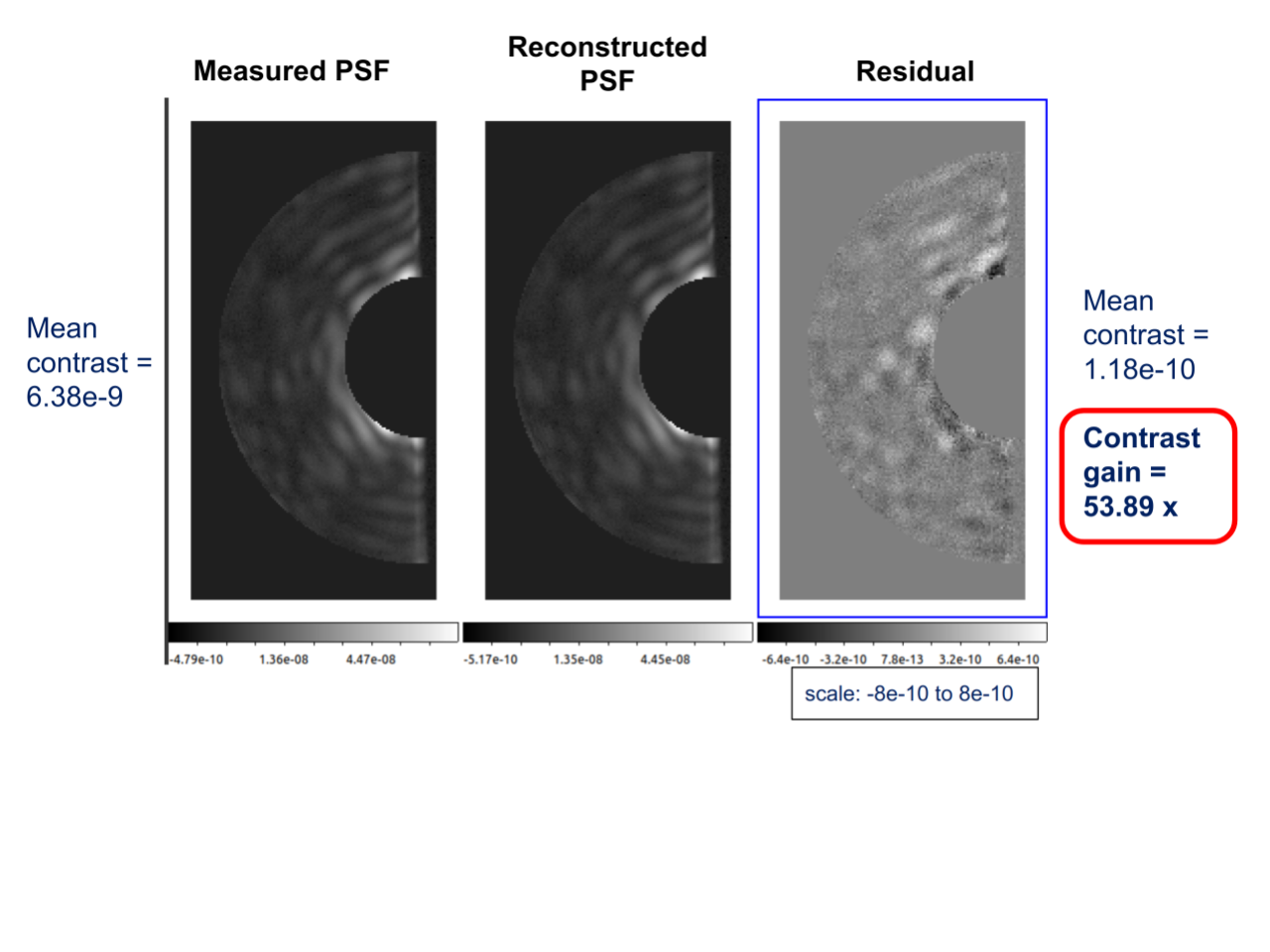}

    \vspace{-0.4in}
    \caption{Vacuum test of sLDFC-based PSF reconstruction with the JPL High-Contrast Imaging Testbed \citep{Guyon2024}.   Spatial LDFC signals are used here to reconstruct the PSF.  The residuals between the measured DH state and the sLDFC-reconstructed DH are on the order of 10$^{-10}$, implying a contrast gain of over a factor of 50.
    }
\end{figure}

%% Page 4: Observation Description, 1-page limit for this section, 1 page limit for this section
\textbf{\Large Observing Description:}
The program is generally not demanding on target star properties.  It requires a very bright ($V$ = 1--3) PSF reference for DM probing and initial one-sided DH construction.   We then need a very bright to moderately bright ($V$ = 1--5) target star for which the DH-generating DM shape is applied.  At this point, the sLDFC WFC loop is activated and the DH state frozen by sensing the target star's bright field region.   To provide a close parallel to realistic CGI TTR5 and science observations, we follow the ``standard typical observing sequence" of the July 2025 Roman Coronagraph slide deck to allow angular differential imaging and reference star differential imaging (ADI, RDI)\footnote{Coronagraph$\_$CPP$\_$WPoverview2025$\_$8July2025}.
%Each "visit" consists of dark hole digging on a bright nearby PSF reference, PSF reference observations, two sets of +/- telescope rolls ($\Delta$$\theta$= +/- 15$^{o}$) on the target star, followed by a second set of PSF reference observations.  
We will compare the contrast and spatial correlation of the DH speckles over the target sequence to the initial DH state generated by the first PSF reference observation and to subsequent PSF reference star visits.  We will then repeat the experiment and analysis without sLDFC to assess the gain from sLDFC.  

The chief challenge with using sLDFC on CGI is high dynamic range: the Roman Coronagraph's EMCCD is very sensitive to high flux in the bright region when we need sufficiently long exposure times/high gain to illuminate the dark hole speckle floor.  The bright field could easily be too bright to support standard photon counting mode, while analog mode results in substantial ($\sim$ 160 e-). read noise.

Possible ways to perform sLDFC in photon counting mode include but are not limited to: 1) running LDFC on a subset of the dark hole region where the dynamic range is less challenging (e.g. $\sim$ 6--9 $\lambda$/D instead of 3—9 $\lambda$/D), 2) running sLDFC in combination with the wide-field shaped-pupil coronagraph, which has a larger IWA, or 3) using DM probing to reduce the average halo by a factor of 5--10 (and still in the linear response regime) before switching to probing that digs a dark hole on one side.
%Spatial LDFC has enormous potential to enhance the technological and scientific return of the Roman coronagraph. 
While sLDFC only allows a one-sided DH on Roman, current leading detectable Roman Coronagraph targets with known positions will be the focus \citep[e.g.][]{Greco2015,Currie2023a,Currie2025,ElMorsy2025}.  Nothing is lost by implementing sLDFC aside from a small finite-element contrast penalty due to a smaller number of speckle realizations at a given $\lambda$/D \citep{Mawet2014}.

%The rollfollow exactly what is baselined to fulfill TTR5
%\textcolor{blue}{Please include a description (up to 1 page) of the planned observing idea, with an estimate of the observing time necessary to achieve your goals.  This is the portion of the white paper where you can include details on: the choice of the coronagraph mask and bandpass to be used, required exposure sequence, required detection limit and detection S/N, number and spacing of visits, whether angular differential imaging (ADI) or reference differential imaging (RDI) or both are needed, as appropriate for your particular observing idea.}

\textbf{Estimate of Time Needed}:  While no specific target is required, we use Corgi-ETC to estimate the program time for a fidicial V = 4.9 star at 40 $pc$ (e.g. like HIP 99770 \citep{Currie2023a}) assuming a solar system-like exozodi level.  To achieve a 10-$\sigma$ contrast of 2.5$\times$10$^{-8}$ on this star at a small angular separation of 0\farcs{}225 requires 0.64 hrs of integration time for CGI's optimistic performance or 6.4 hours for the conservative case.  These numbers translate into roughly 1.1 hrs and 11 hrs wall time.  At 0\farcs{}4, these numbers reduce to 0.95 hours and 3.33 hours.  Considering the control experiment (no sLDFC) and assuming conservative performance, this program requires at most total of 22 hours at 575 nm with the HLC.  For the wide field imaging with the shaped pupil coronagraph, aiming for a slighlty more modest contrast of 5 $\times$10$^{-8}$ at 0\farcs{}45 results in a similar program time (1.2--2.9 hours integration time and 4--10 hours of wall clock time).
%the required integration time to achieve a 10-$\sigma$ contrast of 2.5 $\times$10$^{-8}$ is roughly a factor of three longer.  

\printbibliography

\end{document}